\pdfoutput=1 
\documentclass[english,11pt,a4paper]{article}

\usepackage{jheppub}
\usepackage{graphicx}
\usepackage{bm}
\usepackage{amsmath}
\usepackage{lineno}
\usepackage{amssymb}
\usepackage[utf8]{inputenc}
\usepackage{lipsum}
\usepackage{stmaryrd}

\IfFileExists{dsfont.sty}
	{\usepackage{dsfont}
         \let\mathbb=\mathds
         \newcommand{\id}{\mathds{1}}}
	{\typeout{Package dsfont.sty was not found, using alternative macros.}
         \let\mathds=\mathbb
         \newcommand{\id}{\mbox{1 \kern-.59em {\rm l}}}}

\renewcommand\a{\alpha}
\renewcommand\b{\beta}
\renewcommand\d{\delta}

\renewcommand\l{\lambda}
\renewcommand\r{\rho}

\renewcommand\o{\omega}

\newcommand\e{\epsilon}
\newcommand\g{\gamma}

\newcommand\m{\mu}
\newcommand\n{\nu}
\newcommand\x{\xi}
\newcommand\p{\pi}

\newcommand\s{\sigma}

\renewcommand\L{\Lambda}

\renewcommand\O{\Omega}

\newcommand{\bO}{{\bm \O}}

\newcommand{\bB}{{\bm B}}
\newcommand{\bE}{{\bm E}}
\newcommand{\bJ}{{\bm J}}

\renewcommand{\vec}{\boldsymbol}
\newcommand{\vx}{\vec x}
\renewcommand{\part}{{\rm part}}

\newcommand{\be}{\begin{equation}}
\newcommand{\ee}{\end{equation}}
\newcommand{\bes}{\begin{subequations}}
\newcommand{\ees}{\end{subequations}}
\newcommand{\bea}{\begin{eqnarray}}
\newcommand{\eea}{\end{eqnarray}}

\newcommand{\pa}{\partial}

\newcommand{\na}{\nabla}

\newcommand{\lrd}{\overset{\leftrightarrow}{\partial}}

\def\nbox#1#2{\vcenter{\hrule \hbox{\vrule height#2in
\kern#1in \vrule} \hrule}}
\def\sq{\,\raise.5pt\hbox{$\nbox{.10}{.10}$}\,}
\def\sqb{\,\raise.5pt\hbox{$\overline{\nbox{.09}{.09}}$}\,}

\allowdisplaybreaks
\begin{document}

\title{Zilch Vortical Effect, Berry Phase, and Kinetic Theory}

\author[a,b]{Xu-Guang Huang,}
\author[c,d]{Pavel Mitkin,}
\author[c,e]{Andrey V. Sadofyev,}
\author[f]{and Enrico Speranza}

\affiliation[a]{Physics Department and Center for Particle Physics and
             Field Theory, Fudan University, Shanghai 200433, China}
\affiliation[b]{Key Laboratory of Nuclear Physics and Ion-beam Application(MOE), Fudan University, Shanghai 200433, China}
\affiliation[c]{ITEP, B. Cheremushkinskaya 25, Moscow 117218, Russia}
\affiliation[d]{Moscow Institute of Physics and Technology, Dolgoprudny, Moscow 141700, Russia}
\affiliation[e]{Theoretical Division, MS B283, Los Alamos National Laboratory, Los Alamos, NM 87545, USA}
\affiliation[f]{Institute for Theoretical Physics, Goethe University, Max-von-Laue-Str. 1, D-60438 Frankfurt am Main, Germany}

\emailAdd{huangxuguang@fudan.edu.cn, pavel.mitkin@phystech.edu, sadofyev@lanl.gov, esperanza@itp.uni-frankfurt.de}

\abstract{
Rotating photon gas exhibits a chirality separation along the angular velocity which is manifested through a generation of helicity and zilch currents. In this paper we study this system using the corresponding Wigner function and construct elements of the covariant chiral kinetic theory for photons from first principles. The Wigner function is solved order-by-order in $\hbar$ and the unconstrained terms are fixed by matching with quantum field theory results.  We further consider the zilch and helicity currents and show that  both  manifestations of the chirality transport originate in the Berry phase of photons similarly to other chiral effects. Constructing the kinetic description from the Wigner function we find that the frame vector needed to fix the definition of spin of a massless particle is, in fact, the vector of the residual gauge freedom for the free Maxwell theory. We also briefly comment on the possible relation between vortical responses in rotating systems of massless particles and the anomalies of underlying quantum field theory.
}

\date{\today}
\maketitle

\section{Introduction}
\label{Sec:Introduction}
Systems of chiral particles can exhibit a set of novel transport phenomena closely tied to the anomalies of the underlying quantum field theory (QFT). These chiral effects can considerably influence dynamics of a variety of systems from quark-gluon plasma (QGP) in heavy-ion collisions to Weyl and Dirac semi-metals and attracted a lot of attention in the literature, for review see \cite{Kharzeev:2015znc, Huang:2015oca}.

Recently, it was suggested that some chiral effects can take place for photons and other massless particles and quasi-particles with spin $s>\frac{1}{2}$, see e.g. \cite{Huang:2018aly, Avkhadiev:2017fxj}. Particularly, in a rotating gas of photons one may expect a separation of circular polarizations along the angular velocity -- the chiral vortical effect (CVE) of photons \cite{Avkhadiev:2017fxj, Yamamoto:2017uul, zuyizin, Prokhorov:2020okl}. That generates a non-zero magnetic helicity current of photons which is also related to the corrections to the fermionic CVE in the axial current \cite{Hou:2012xg, Golkar:2012kb}. This separation may affect the common dynamics of microscopic and macroscopic helicities in chiral media leading to a new class of instabilities, see e.g. \cite{Akamatsu:2013pjd, Khaidukov:2013sja, Kirilin:2013fqa, Avdoshkin:2014gpa, Manuel:2015zpa, Buividovich:2015jfa, Yamamoto:2015gzz, Hirono:2015rla, Kirilin:2017tdh, Li:2017jwv, Tuchin:2019gkg, Mace:2019cqo, Horvath:2019dvl}.  However, while the corresponding total helical charge is well-defined, the local helicity current 
\be
K^\mu=\epsilon^{\mu\nu\a\b}A_\nu F_{\a\b}
\ee
is gauge dependent.
This issue can be addressed if one finds a gauge-invariant and local choice for the measure of the photon polarization transport.

The non-interacting Maxwell theory does have an infinite set of extra conserved gauge-invariant currents sensitive to the polarization transport and known as zilches \cite{lipkin, kibble}.
The corresponding zilch charges count the difference between the number of right- and left-handed photons  weighted with the even power of their energy and summed over the phase space. In \cite{cohen2010} it was shown that the lowest zilch determines the chiral asymmetry in the response of a chiral molecule to external electromagnetic (EM) fields  and one can take it as a gauge-invariant and local measure of the optical chirality. The general zilch current possesses a contribution similar to the photonic CVE -- the zilch vortical effect (ZVE), see \cite{Chernodub:2018era, Copetti:2018mxw}. However, the relation between CVE and ZVE in systems of photons has not been studied in the literature in details.

While the chiral effects in external electromagnetic fields are explicitly connected to the axial anomaly, the microscopic origin of the chiral effects in rotating systems is still under intense discussion, see e.g. \cite{Kharzeev:2015znc, Huang:2015oca} and references therein. Indeed, vortical effects take place even in the absence of EM fields when the axial anomaly is turned off. Moreover, the CVE in the axial current of fermions at finite temperature is present in the limit of zero charge of constituents. This portion of the fermionic CVE is suggested to originate in the mixed gravitational anomaly \cite{Landsteiner:2011cp, Landsteiner:2011iq} or global gravitational anomaly \cite{Hou:2012xg, Golkar:2012kb}, although these conjectures are still under discussions, see e.g. \cite{Avkhadiev:2017fxj, Prokhorov:2020okl, Glorioso:2017lcn, Flachi:2017vlp, Stone:2018zel}. If the relation between the thermal part of CVE and the mixed gravitational anomaly is used as a guidance \cite{Avkhadiev:2017fxj, Chernodub:2018era}, the photonic CVE is also expected and can be attributed to the axial anomaly of photons in an external gravitational field \cite{Dolgov:1987yp, Vainshtein:1988ww, Dolgov:1988qx, Agullo:2016lkj}. However, the foreseen relation between photonic CVE and ZVE is seemingly hard to fit in the same picture unless a new class of anomalies exists for the zilch currents in a background gravitational field.

On the other hand, the origin of ZVE and its relation to CVE can be studied in the semi-classical limit, in the framework of chiral kinetic theory (CKT). Then one may notice that the anomalous chiral transport of fermions shares similarities with the spin Hall effect and originates in the topological Berry phase. This relation between the fermionic anomalous chiral transport and the Berry phase was first observed in terms of single particle semi-classical action \cite{Son:2012wh, Stephanov:2012ki, Son:2012zy} and then generalized to chiral particles of arbitrary spin \cite{Huang:2018aly, Yamamoto:2017uul, Yamamoto:2017gla} within the same approach. Thus, one may ask if ZVE has the same origin as CVE and is related to topological phases arising in the CKT description allowing further discussion on their anomalous nature. However, writing field-theoretical objects in the CKT terms requires one to closely follow how the CKT appears from the underlying microscopic theory. Such an identification between the kinetic theory and QFT attracted significant attention recently. Using the Wigner-function formalism it was shown that the fermionic CKT indeed follows from the underlying theory and the current expectation values were identified in the CKT terms correctly \cite{Chen:2012ca, Gao:2012ix,Hidaka:2016yjf,Huang:2018wdl,Gao:2018wmr,Prokhorov:2018qhq,Liu:2018xip}.

In this work, we utilize the same tools and show that the CKT of photons also arises from the Wigner-function formalism. We consider a simple field theoretical calculation for ZVE closely following \cite{Chernodub:2018era} and identify a possible form of the general zilch current in the semi-classical description. Then we introduce a ``naive'' CKT for photons and show that the zilch current written in CKT terms gains the same ZVE contribution in this limit. Strikingly, we find that the ZVE originates in the non-trivial topological Berry phase and, in this sense, is similar to the CVE. We further review the basics of the Wigner function description for gauge fields \cite{Vasak:1987um, Elze:1986hq,Elze:1989un} and construct the Wigner function for photons in a rotating system. The equations of motion (EOMs) of the Wigner function cannot fix it completely and we utilize the QFT results to remove the arbitrariness. Using this Wigner function we show that the CKT definition of the zilch current indeed corresponds to its QFT definition and, in this way, support our CKT construction. Thus, we identify the relation between vortical effects of photons in different observables and provide a basis for further studies of the gauge field contribution to spin polarization phenomena.

Throughout this work we use the most-negative signature for the spacetime metric and the convention $\e^{0123}=-\e_{0123}=1$ for the Levi-Civita symbol.

\section{Zilch currents and rotating thermal radiation}
\label{Sec:QFT}
Let us start with a brief discussion of the zilch current definition and its properties \cite{lipkin, kibble}. This current and the corresponding charge are originally introduced in \cite{lipkin} as non-covariant 3-dimensional objects constructed out of electromagnetic fields
\bea
\bJ_Z=\frac{1}{2}\left(\bE\times\dot{\bE}+\bB\times\dot{\bB}\right)\,,~~~~Z=\frac{1}{2}\left(\bB\cdot \dot{\bE}-\bE\cdot\dot{\bB}\right)\,,
\eea
which are conserved on the free Maxwell equations
\bea
\pa_0 Z+\vec{\na}\cdot\bJ_Z=0\,.
\label{JZcons}
\eea
The normalization is chosen in such a way that the corresponding charge $Z$ involves differences between number of right- and left-handed circularly polarized photons weighted with photon energy squared when the theory is quantized~\cite{Chernodub:2018era}. One can notice that the zilch charge and current are actually components of a rank-3 tensor
\bea
Z^{(3)}_{\mu\nu\r}=\frac{1}{2}\left(\tilde{F}_\mu^{~\l}\pa_\r F_{\l\nu}-F_\mu^{~\l}\pa_\r \tilde{F}_{\l\nu}\right)\,,
\label{zilch_tensor_old}
\eea
where $\tilde{F}^{\m\n}=\frac{1}{2}\epsilon^{\m\n\r\s}F_{\r\s}$ and the conservation (\ref{JZcons}) is equivalent to $\pa^\m Z^{(3)}_{\m00}=0$.  As shown in \cite{kibble}, in fact, one can introduce an infinite tower of higher-rank tensorial zilch currents adding derivatives to the definition above. Then the general zilch current is given by
\bea
Z^{(s)}_{\a_1...\a_s}=\frac{1}{2}\left(\tilde{F}_{\a_1}^{~\l}\pa_{\a_2}..\pa_{\a_{s-1}} F_{\l\a_s}-F_{\a_1}^{~\l}\pa_{\a_2}...\pa_{\a_{s-1}} \tilde{F}_{\l\a_s}\right)
\label{zilch_old_s}
\eea
and satisfies $\pa^{\a_1}Z^{(s)}_{\a_1...\a_s}=0$. Combining these zilch currents with other conserved quantities, a modified set of conserved currents (with the same charges) can be constructed and, thus, the general zilch current is defined not uniquely.

In a large cylinder rotating with angular velocity $\vec \O$ and being at equilibrium with thermal radiation of photons one may expect a polarization transport due to photonic CVE, see \cite{Avkhadiev:2017fxj, Yamamoto:2017uul, Huang:2018aly}. In this system the zilch current also has a non-zero expectation value \cite{Chernodub:2018era} giving a gauge-invariant measure of the chirality transport
\bea
{Z^{(3)i}}_{00}=\frac{8\pi^2 T^4}{45}\O^i\,.
\label{zilchKarl}
\eea
However, in this work we are interested in studying ZVE in CKT and it will be more convenient to use a zilch definition with higher degree of symmetry allowing to identify the corresponding object in the single-particle language. A direct check shows that the fully symmetric Lorentz tensor obtained from the zilch current is also conserved, see e.g. \cite{Copetti:2018mxw}. Thus, one may introduce another set of conserved zilch currents $\pa^{\a_1}\bar{Z}^{(s)}_{\a_1..\a_s}=0$ given by
\bea
\bar{Z}^{(s)}_{\a_1..\a_s}=\tilde{F}_{\l\{\a_1}\lrd_{\a_2}...\lrd_{\a_{s-1}} F_{\a_s\}}^{~~~\l}\,,
\label{zilch_new_s}
\eea
where we introduce an index symmetrization $A_{\{\a_1...\a_s\}}=\frac{1}{s!}\sum A_{\Pi(\a_1...\a_s)}$ with the sum going over all permutations $\Pi$ and two-way derivative $\lrd=\frac{1}{2}\big(\overset{\rightarrow}{\partial}-\overset{\leftarrow}{\partial}\big)$. Notice here that the non-trivial zilch currents must have odd ranks so $s=2k+1$ with $k\in\mathbb{Z}$ and we will assume that in what follows.

Using the definition (\ref{zilch_new_s}) we can derive the expectation value of the general symmetric zilch current $\left\langle \bar{Z}^{(s)}_{\a_1..\a_s}\right\rangle$ in a rotating thermal radiation closely following \cite{Chernodub:2018era}. The equilibrium expectation value of an operator is given by
\be
\left\langle O(\vx)  \right\rangle = \text{Tr}\left[ \r\,O(\vx, t) \right]
\label{Oexp}
\ee
where $\r = C \exp \left[ -\b \left( H - \bm M\cdot\bO \right) \right]$ is the statistical operator in a rotating system \cite{Vilenkin:1980zv}, $H$ is the Hamiltonian, $\bm M$ is the angular momentum operator, $C$ is a normalization constant, and $\b=T^{-1}$. Without loss of generality we take $\O$ along the $z$-direction and work in the cylindrical coordinates $x=(r,\phi,z)$. We take the Coulomb gauge $\frac{1}{\sqrt{-g}}\partial_i\sqrt{-g}\vec{A}^i=0$ which  additionally results in $A_0=0$ in the absence of sources in the Maxwell equations. Solving the Maxwell equations we use that the system is invariant under translations in time, angle, and along the cylinder axis with an elementary solution obeying $A^i_{\l\o k m}\propto e^{-i\o t+i k z+i m\phi}$, where $\l=\pm$ corresponds to the two possible polarizations, $\o$ is the frequency, $m\in\mathbb{Z}$ is the angular quantum number, and $k$ is the linear momentum along the $z$-direction.

We are interested in the polarization transfer and choose the basis of circularly polarized waves given by the condition
\bea
\tilde{F}^{\m\n}_{\pm,\o k m}\pm iF^{\m\n}_{\pm,\o k m}=0\,,
\label{PolCond}
\eea
where $\tilde{F}^{\m\n}=\frac{1}{2\sqrt{-g}}\e^{\m\n\r\s}F_{\r\s}$ for the curvilinear coordinates with $\e^{\m\n\r\s}$ the usual Levi-Civita symbol. In the Coulomb gauge the Maxwell equations reduce to a cylindrical wave equation and can be directly solved for. One can explicitly check that the harmonics satisfying (\ref{PolCond}) and the Maxwell equations are given in terms of Bessel functions and read
\bea
\vec{A}_{\pm\o k m}
=\frac{1}{\sqrt{2}k_\perp}
\left(\begin{array}{c}
-\frac{im}{r} J_m (k_\perp r )\mp\frac{i k}{\o}\frac{\pa}{\pa r}J_m(k_\perp r)  \\
\frac{1}{r}\frac{\pa}{\pa r}J_m (k_\perp r) \pm \frac{mk}{\o} \frac{J_m(k_\perp r)}{r^2} \\
\mp\frac{k_\perp^2}{\o} J_m(k_\perp r)
\end{array}\right)e^{-i\o t+i k z+i m\phi}\,,
\label{harmonic}
\eea
where $k_\perp^2=\o^2-k^2$ is the transverse/radial momentum, the subscript corresponds to the sign in (\ref{PolCond}), and the overall factor is chosen in the standard way to lead to canonically-normalized quantization relation. The boundary conditions may result in quantization of $k_\perp$ but we are interested in the limit of an unbounded space and $k_\perp$ is continuous. As usual the full set of harmonics form a complete orthonormal basis
\bea
\int\,d^3x\,\sqrt{-g}\,g_{ij}\,A^i_{\l\o k m}(t,\vec{x})\,A^{j,\star}_{\l'\o' k' m'}(t,\vec{x})=-4\p^2\d_{\l\l'}\d_{mm'}\delta(k-k')\frac{\delta(k_\perp-k'_\perp)}{k_\perp}\,,
\eea
where the delta functions on the r.h.s. correspond to the choice of cylindric coordinates in the momentum space. Thus, using the general solution of the Maxwell equations we can write the photon field operator as
\bea
A^i=\sum_{\l, m}\int_0^\infty\,\frac{k_\perp dk_\perp}{2\p}\,\int_{-\infty}^\infty\,\frac{dk}{2\p}\,\frac{1}{\sqrt{2\o}}\left[a_{\l\o k m}A^i_{\l\o k m}(t,\vec{x})+a^\dag_{\l\o k m}A^{i,\star}_{\l\o k m}(t,\vec{x})\right]\,,
\eea
where, as usual, $a_{\l\o k m}$ and $a^\dag_{\l\o k m}$ are the creation and annihilation operators satisfying the canonical commutation relation
\bea
\left[a_{\l\o k m},a^\dag_{\l^\prime\o^\prime k^\prime m^\prime}\right]=4\p^2\d_{\l\l^\prime}\d_{mm^\prime}\d(k-k^\prime)\frac{\d(k_\perp-k^\prime_\perp)}{k_\perp}\,.\notag
\eea
Now we can calculate the expectation value $\left\langle \bar{Z}^{(s)}_{\a_1..\a_s}\right\rangle$ in a rotating system in the thermal equilibrium. If the vacuum is defined in the non-rotating laboratory frame (for a static observer), then applying (\ref{Oexp}) we find
\bea
\label{expect}
\langle a^\dag_{\l^\prime\o^\prime k^\prime m^\prime} a_{\l\o k m}\rangle=4\p^2\,f_B(\x)\,\d_{\l\l^\prime}\d_{mm'}\delta(k-k')\frac{\delta(k_\perp-k'_\perp)}{k_\perp}\,,~~~
\eea
where $f_B(\x)=\left[e^{\x}-1\right]^{-1}$ is the Bose-Einstein distribution function for a rotating ensemble and $\x=\b\left(\o-m\O\right)$ in cylindrical coordinates for a static observer. However, a rotating relativistic system cannot be unbounded and one has to require that $\O R<1$. Thus, the singularity in the distribution function must be regularized by the finite size effects on the soft modes with $k_\perp=\L_{IR}\sim \O$. Following \cite{Vilenkin:1980zv} we assume that the finite-size effects are suppressed in the limit $T\gg\frac{1}{R}$ and focus on the leading contributions in the powers of the temperature, for an additional discussion see \cite{Chernodub:2018era}. The expectation value of the normal ordered zilch current can be expressed as 
\bea
\label{zilchsymm}
\left\langle :\bar{Z}^{(s)}_{\a_1..\a_s}(\vx, 0):\right\rangle=\sum_{\l, m}\int_{\L_{IR}}^\infty\,\frac{k_\perp dk_\perp}{2\p}\,\int_{-\infty}^\infty\,\frac{dk}{2\p}\,f_B(\x)\,z^{\l\o k m}_{\a_1..\a_s}\,,
\eea
where $z^{\l\o k m}_{\a_1..\a_s}$ is the value of (\ref{zilch_new_s}) on a single harmonic (\ref{harmonic}). The zilch current is a Lorentz $s$-tensor and its all-temporal-but-one-spacial component in the laboratory frame gives the desired measure of the photon polarization transfer. After simple algebra one finds
\bea
\sum_\l\,z^{\l\o k m}_{30..0}=(-1)^{\frac{s-1}{2}}\frac{4\,m\,\o^{s-1}}{s}\left(1+(s-1)\frac{k^2}{\o^2}\right)\frac{J_m(k_\perp r)J'_m(k_\perp r)}{k_\perp r}\,,
\eea
and the value of the general rank-$s$ zilch current can be explicitly calculated. On the rotation axis ($r\to0$) this expression is non-zero only for $m=\pm1$ and in the limit of $\O\to0$ there is no zilch current since the the two contributions cancel. We are interested in the leading linear term in $\O$ which reads
\bea
&&\left\langle :\bar{Z}^{(s)}_{30..0}:\right\rangle\bigg|_{r\to0}= (-1)^{\frac{s-1}{2}}\frac{(s+2)(s+1)}{3s}\frac{\O}{\p^2}\int_0^\infty\,d\o\,\o^{s}\,f_B(\b\o)+\mathcal{O}(\O^2)\,,
\label{eq:ZilchQFT}
\eea
giving the value of ZVE in the general symmetrized zilch current on the rotation axis for a static observer, c.f. \cite{Chernodub:2018era}.\footnote{We also notice here that our result for ZVE in the symmetrized zilch current obtained in QFT disagrees with \cite{Copetti:2018mxw}. However, as we will see (\ref{eq:ZilchQFT}) is reproduced in the Wigner-function formalism along with the correct ZVE in the non-symmetrized zilch current (\ref{ZVEKarl}).}

\section{Chiral Kinetic Theory}
\label{Sec:CKT}
Let us now turn to a covariant formulation of CKT introduced in \cite{Chen:2015gta} and briefly discuss the issues with the Lorentz properties of this theory. The fermionic case considered in \cite{Chen:2015gta} can be readily extended to the general case of a massless particle with an arbitrary spin \cite{Huang:2018aly}. The decomposition of the full angular momentum of a relativistic massless particle
\bea
J^{\m\n}=x^\m p^\n-x^\n p^\m+S^{\m\n}
\eea
is ambiguous due to an arbitrariness in the definition of the spin part $S^{\m\nu}$. To fix this ambiguity one may introduce a frame vector $n^\m$ and require that $p_\m S^{\m\n}=n_\m S^{\m\n}=0$ with $(n\cdot p)\neq0$ constraining the spin tensor to have only spatial components in the particular frame. Then $S^{\m\n}$ is uniquely fixed and reads
\bea
S^{\m\n}_n=\l\,\hbar\,\frac{\e^{\m\n\r\s}p_\r n_\s}{p\cdot n}\,,
\eea
where $\l$ is the particle helicity equal to $\pm1$ in the case of photons. Here and in the rest of the text we restore the powers of $\hbar$ working with the semi-classical expansion.

The physical quantities are expected to be independent of the frame choice. Thus, changing the frame vector one has to modify the definition of the coordinate $x$ to keep the full angular momentum unchanged. This shift of the coordinate is nothing else but the side jump effect \cite{Skagerstam:1992er, Chen:2014cla, Duval:2014ppa, Chen:2015gta} -- an additional shift of the coordinate of a massless particle under a frame change at first order in $\hbar$. Due to this shift the distribution function $f$ is frame dependent and the naive current $p^\m f$ is not a Lorentz vector. However, it was shown in \cite{Chen:2015gta} that
\bea
j^\m=p^\m f+S^{\m\n}_n\pa_\n f
\label{CKTj}
\eea
is a 4-vector since the frame dependent shift in $f$ cancels by the modification of $S^{\m\n}_n$.

Now we easily construct a tensor with the right dimensionality and Lorentz properties of the general zilch current density in the phase space
\bea
z^{(s,\l)}_{\a_1..\a_s}=\l(-1)^{\frac{s+1}{2}}p_{\{\a_1}p_{\a_2}...j_{\a_s\}}\,,
\label{CKTzilchSP}
\eea
where the normalization is chosen to match with the zilch defined in QFT. In equilibrium, a system of photons is described by $f(\x)$ with $\x=\b_\m p^\m+\frac{1}{2}S^{\m\n}_n\O_{\m\n}$ in Cartesian coordinates, where $\b^\m$ is the temperature 4-vector satisfying $T^2\b^\m\b_\m=1$ and $\O^{\m\n}=\frac{1}{2}\left(\pa^\m\b^\n-\pa^\n\b^\m\right)$ is the thermal vorticity tensor, see e.g. \cite{Chen:2015gta}. Thus, the CKT zilch current in the laboratory frame is given by a phase space integral of the single particle zilch (\ref{CKTzilchSP}) and reads
\bea
\bar{Z}^{(s)}_{30..0}=\frac{1}{\hbar^{s}}\sum_\l\l(-1)^{\frac{s+1}{2}}\int\frac{d^4p}{(2\pi)^3}\,\d(p^2)\,\,p_{\{0}p_{0}...j_{3\}}\,,
\eea
where we impose the on-shell condition and restore powers of $\hbar$ in the definition of the zilch. The equilibrium current is $n$-independent and, using the Schouten identity, we get the leading terms in the $\hbar$ expansion
\bea
j^\m=p^\m f(\b_\n p^\n)-\frac{1}{2}\l\,\hbar\, \e^{\m\n\r\s}p_\n\O_{\r\s} f'(\b\cdot p)+\mathcal{O}\left(\hbar^2\right)\,,
\eea
where the distribution function of photon gas is $f(\b\cdot p)=\theta(\b\cdot p)f_B(\b\cdot p)+\theta(-\b\cdot p)f_B(-\b\cdot p)$ and a term proportional to $p^2$ is neglected since it does not contribute to the zilch current. Finally, we find that on the rotation axis in the laboratory frame the general zilch current is given by
\bea
\bar{Z}^{(s)}_{30..0}\bigg|_{r\to0}=(-1)^{\frac{s-1}{2}}\frac{(s+2)(s+1)}{3s\hbar^{s-1}}\frac{\O}{\pi^2}\int\,d\o\,\o^s f_B(\b \o)+\mathcal{O}(\O^2)\,,
\label{eq:ZilchCKT}
\eea
where we used the fact that the only non-zero components of $\O_{\m\n}$ are $\O_{21}=-\O_{12} \equiv \frac{\O}{T}$ and $\b=\frac{1}{T}$. Thus, one finds that the symmetric general zilch current in a rotating system derived in CKT agrees with the field theory result. Moreover, in the CKT formulation it is related to the topological Berry phase through the side-jump term in the current and the spin-orbit coupling in the distribution function which are proportional to $S^{\m\n}_n$. This is one of the main results of this paper.

\section{Wigner function for photons}
\label{Sec:Wigner}

The Wigner-function formalism has been widely used to construct quantum kinetic theory, see e.g. \cite{Chen:2012ca, Gao:2012ix,Hidaka:2016yjf,Huang:2018wdl,Gao:2018wmr,Gao:2019znl,Prokhorov:2018qhq,Liu:2018xip, Hattori:2019ahi,Wang:2019moi,Weickgenannt:2019dks,Liu:2020flb}. This approach is particularly convenient to study spin-polarization phenomena due to system rotation and external EM fields, and allows us to describe  non-equilibrium effects. However, most recent considerations are focused solely on the fermionic degrees of freedom. In this section we will follow the early works \cite{Vasak:1987um, Elze:1986hq,Elze:1989un} and construct the Wigner function for photons in a rotating system. Using this Wigner function, the CKT definition of the zilch current can be obtained directly. It will also allow us to re-express the relation of ZVE and CVE to the Berry phase in terms of the underlying many-body QFT and  connect the CKT and QFT calculations of the previous sections.

We start by defining a gauge-dependent Wigner function of spin-one Abelian gauge fields
\begin{equation}
W^{\mu\nu}(x,p)=\int \frac{d^4y}{(2\pi \hbar)^4} e^{-\frac i \hbar p\cdot y}  \langle : A^{\mu}\Big(x+\frac{y}2\Big) A^{\nu}\Big(x-\frac{y}2\Big) : \rangle\,,
\label{Wdef}
\end{equation}
where the Wigner representation is used. If the interactions are ignored the Wigner function satisfies the same EOMs as the free fields and in the Lorenz gauge $\partial_\mu A^\mu =0$ they read
\bea
\Bigg(p^2-\frac{\hbar^2}{4}\partial^2 \Bigg)W^{\mu\nu}(x,p) &=& 0 \, ,\\
\hbar \, p\cdot \partial W^{\mu\nu}(x,p) &=& 0\,.
\eea
To satisfy the gauge constraint we also require that
\bea
\Bigg(p_\alpha -i\frac\hbar2 \partial_\alpha\Bigg)W^{\alpha\mu}(x,p)=\Bigg(p_\alpha +i\frac\hbar2 \partial_\alpha\Bigg)W^{\mu\alpha}(x,p)=0
\eea
and fix the residual gauge freedom for free Maxwell theory with
\bea
n_\alpha W^{\alpha\mu}(x,p)=n_\alpha W^{\mu\alpha}(x,p)  =0\,,
\eea
where $n^\mu$ is taken to be a constant time-like vector satisfying $n^2=1$. Notice that for free EM fields the Lorenz gauge condition is, for instance, fully compatible with the Coulomb gauge and $n^\m$ is introduced to classify a family of such gauges.

We study these equations by employing an expansion in powers of $\hbar$ which is equivalent to a gradient expansion. Hence, we search for a solution of the type
\begin{equation}
W^{\mu\nu}=W^{(0)\mu\nu}+ \hbar W^{(1)\mu\nu}+\dots \,.
\end{equation}
The leading order EOMs take the form
\bea
&&p^2W^{(0)\mu\nu}(x,p) = p\cdot \partial W^{(0)\mu\nu}(x,p)  =0\,.
\eea
Additionally, enforcing the gauge-fixing constraints we have to require
\bea
&&p_\alpha W^{(0)\alpha\mu}(x,p) =p_\alpha W^{(0)\mu\alpha}(x,p)=n_\alpha W^{\alpha\mu}(x,p)=n_\alpha W^{\mu\alpha}(x,p)  =0\,.
\eea
To proceed further one has to assume the form of the leading order Wigner function $W^{(0)}$ which is not uniquely determined by the EOMs and constraints. Here we use an ansatz motivated by a field theory calculation of (\ref{Wdef}) for free fields in the case of a static uniform gas
\begin{equation}
\label{ord0}
W^{(0)\mu\nu}(x,p)=P_n^{\mu\nu} F(x,p)\delta (p^2)\, ,
\end{equation}
where $P_n^{\mu\nu}=-g^{\mu\nu}+\frac{p^\m n^\n+p^\n n^\m}{p\cdot n}-\frac{p^\mu p^\nu}{(p\cdot n)^2}$ is an on-shell gauge projection operator and the unspecified distribution function $F(x,p)$ satisfies the Boltzmann equation $p \cdot \partial F(x,p)=0$. Note that $W^{(0)\mu\nu}$ in general does not need to be symmetric but we expect the anti-symmetric terms to go to zero in the uniform limit. At first order in $\hbar$, the EOMs read
\bea
p^2W^{(1)\mu\nu}(x,p) = p\cdot \partial W^{(1)\mu\nu}(x,p) =0,
\eea
with $W^{(1)\m\n}$ additionally constrained by
\bea
 p_\alpha W^{(1)\alpha\mu}(x,p) - \frac{i}{2} \partial_\alpha W^{(0)\alpha\mu}(x,p)&=0 \, ,\\
 p_\alpha W^{(1)\mu\alpha}(x,p) +\frac{i}{2} \partial_\alpha W^{(0)\mu\alpha}(x,p) &=0 \, , \\
n_\a W^{(1)\a\m}(x,p) = n_\a W^{(1)\m\a}(x,p)  &=0 \, .
\label{nW1}
\eea
In order to construct the general $W^{(1)\mu\nu}$ we first write the Wigner function as a sum of a symmetric and antisymmetric part $W^{(1)\mu\nu}=W^{(1)\mu\nu}_S+W^{(1)\mu\nu}_A$. Then the Lorenz gauge constraints are equivalent to
\bea
p_\alpha W^{(1)\alpha\mu}_S &= 0\, ,~~~~~~p_\alpha W^{(1)\alpha\mu}_A &= \frac{i}2 P_n^{\m\a}\partial_\a  F(x,p)  \delta (p^2)\,.
\label{antisym1}
\eea
As we will discuss later the symmetric part of the Wigner function cannot contribute to the polarization transport and, thus, we focus solely on $W^{(1)\m\n}_A$. Without loss of generality we can parametrize the antisymmetric part of the Wigner function satisfying (\ref{nW1}) as
\begin{equation}
\label{WAdec}
W^{(1)\mu\nu}_A = \epsilon^{\m\n\r\s}\,n_\r H_\s(x,p)\delta (p^2)\,,
\end{equation}
where $H_\s$ is a generic function such that $H\cdot n=0$. Substituting (\ref{WAdec}) into (\ref{antisym1}) one finds that
\begin{equation}
\epsilon^{\m\n\r\s}\,p_\n n_\r  H_\s(x,p)\delta (p^2)=-\frac{i}2 P_n^{\m\a}\partial_\a  F(x,p)\delta (p^2)
\end{equation}
and, in the general case,
\begin{equation}
H_\m(x,p)=-\frac{i}2 \e_{\m\n\r\s}\,\frac{p^\n n^\r}{(p\cdot n)^2}\partial^\s  F(x,p)-i\frac{\tilde{p}_\m}{(p\cdot n)^2} U(x,p)\,,
\end{equation}
where $\tilde{p}^\m=p^\m-n^\m(p\cdot n)$ is a projection of $p^\m$ transverse to $n^\m$, $U$ is a free function unconstrained by the EOMs, and we have used $p \cdot \partial F(x,p)=0$. Thus, the general solution for $W_A^{(1)}$ can be written as
\begin{equation}
\label{ord1}
W^{(1)\mu\nu}_A = -\frac{i}{2}\frac{\tilde{p}^{[\m} P_n^{\n]\a}}{(p\cdot n)^2} \partial_\alpha F(x,p) \delta (p^2)+i\epsilon^{\m\n\r\s}\frac{p_\r n_\s}{p\cdot n} U(x,p) \delta (p^2)\,,
\end{equation}
where $a^{[\m} b^{\n]}=a^\m b^\n-a^\n b^\m$.

As we will see, the free function $U$ contributes to the zilch current and the Wigner function should be further constrained. However, this problem can be partially resolved if one requires gauge invariance of the Wigner function defined for the field strength tensor
\bea
&&Y^{\m\nu\r\s}(x,p)=\hbar^2\int \frac{d^4y}{(2\pi \hbar)^4} e^{-\frac i \hbar p\cdot y}  \langle:F^{\m\n}\Big(x+\frac{y}2\Big)F^{\r\s}\Big(x-\frac{y}2\Big):\rangle
\eea
which satisfies $\left(p_\m-\frac{i}{2}\hbar\pa_\m\right) Y^{\m\nu\r\s}(x,p)=\left(p_\r+\frac{i}{2}\hbar\pa_\r\right) Y^{\m\nu\r\s}(x,p)=0$. Notice that we rescale $Y^{\m\nu\r\s}$ with powers of $\hbar$ to link the gradient and semi-classical expansions. The zeroth-order contribution to $Y^{\m\nu\r\s}$ can be readily obtained from $(\ref{ord0})$ and is given by an explicitly gauge-invariant expression
\bea
Y^{(0)\m\nu\r\s}(x,p)=-p^{[\m}g^{\n][\s}p^{\r]}F(x,p)\d(p^2)\,.
\eea
Turning to first order, we decompose $Y^{\m\nu\r\s}$ into two parts which are symmetric and antisymmetric under a simultaneous exchange $\m\n\leftrightarrow\r\s$. Similar to $W_S^{(1)}$ we leave $Y^{(1)}_S$ unconstrained since it gives no contribution to the polarization transport and focus on the antisymmetric part
\bea
Y_A^{(1)\m\nu\r\s}(x,p)=p^{[\m}W_A^{(1)\n][\s}p^{\r]}+\frac{i}{2}\pa^{[\r}W^{(0)\s][\n}p^{\m]} -\frac{i}{2}\pa^{[\m}W^{(0)\n][\s}p^{\r]}\,.
\label{Y1def}
\eea
Substituting the gauge-dependent Wigner functions (\ref{ord0}) and (\ref{ord1}) into (\ref{Y1def}) we find
\bea
Y_A^{(1)\m\nu\r\s}(x,p)=\!\! &&-\frac{i}{2}\left(p^{[\m}g^{\n][\s}\pa^{\r]}-p^{[\r}g^{\s][\n}\pa^{\m]}\right)F(x,p)\d(p^2)\notag\\
&&+i\left(\frac{p^{[\m}n^{\n]}p^{[\s}\pa^{\r]}}{p\cdot n}-\frac{p^{[\r}n^{\s]}p^{[\n}\pa^{\m]}}{p\cdot n}\right)F(x,p)\d(p^2)\notag\\
&&+ip^{[\m}\epsilon^{\n]\l\g[\s}p^{\r]}\frac{p_\l n_\g}{p\cdot n} U(x,p) \delta (p^2)\,.~~~~~~~
\label{Ywithn}
\eea
The second and third line in the expression above are explicitly gauge dependent through $n^\mu$ and potentially $U(x,p)$. From now on we assume global equilibrium, set $F=F(\b\cdot p)$, and notice that, on the EOMs, $\b^\m(x)$ satisfies the conformal Killing vector equation $\pa_\m \b_\n+\pa_\n \b_\m= g_{\m\n}\, \phi$ with $\phi$ an arbitrary scalar function, see e.g. \cite{Liu:2018xip}. After some algebra one finds that for $U=\frac{1}{2}\e^{\m\n\r\s}\frac{p_\r n_\s}{p\cdot n}\O_{\m\n} F^\prime(\b\cdot p)+U_0$ we have a gauge-invariant result
\bea
Y_A^{(1)\m\nu\r\s}(x,p)=\!\! &&-\frac{i}{2}\left(p^{[\m}g^{\n][\s}\pa^{\r]}-p^{[\r}g^{\s][\n}\pa^{\m]}\right)F(\b\cdot p)\d(p^2)\notag\\
&&+ip^{[\m}\O^{\n][\s}p^{\r]}F^\prime(\b\cdot p)\d(p^2)+i\epsilon^{\m\n\l[\s}p^{\r]}p_\l U_0(x,p) \delta (p^2)\,,
\label{WF}
\eea
where $U_0$ is $n^\mu$-independent part of $U$.

Using the Wigner function up to first order in $\hbar$ one can derive the leading contribution to the zilch current (\ref{zilch_new_s}) and its expectation value can be written as
\bea
\label{wignerz}
&&\left\langle: \bar{Z}^{(s)}_{\a_1..\a_s}:\right\rangle =2\frac{(-1)^{\frac{s+1}{2}}}{\hbar^{s-1}}\int\,d^4p\,\left[p_{\{\a_2}..p_{\a_{s}}\right]\left(p_{\a_1\}}U+\e_{\a_1\}\m\n\s}\frac{p^\m n^\n}{p\cdot n}\pa^\s F(\b\cdot p)\right)\d(p^2)\,.\notag\\
\eea
This expression exactly agrees with the CKT definition of the zilch current while the expression in the brackets can be related to the CKT current (\ref{CKTj}) introduced in the previous section for $U_0=0$,
\bea
&&\frac{1}{2(2\p)^3\hbar}\left(j_{\a}\big|_{\l=+}-j_{\a}\big|_{\l=-}\right)\simeq p_\a U+\e_{\a\m\n\s}\frac{p^\m n^\n}{p\cdot n}\pa^\s F(\b\cdot p)\,,
\eea
where $F(\b\cdot p)=\frac{1}{(2\p)^3}\left[\theta(\b\cdot p)f_B(\b\cdot p)+\theta(-\b\cdot p)f_B(-\b\cdot p)\right].$
Thus, one may expect that the solution $U_0=0$ is the one corresponding to our setup. Indeed, with this assumption we reproduce the results for ZVE in the general zilch current calculated in the field theory as a function of $s$ matching the Wigner-function formalism to an infinite set of expectation values. The physical meaning of $U$ is one half the difference between distribution functions $f_{\pm}$ for the two polarizations expanded to first order in gradients and, under this assumption, we completely reconstruct the CKT calculation in the Wigner-function formalism. Strikingly, the parameter of the additional gauge symmetry $n^\mu$ plays the role of the frame vector for the spin of photons further clarifying its physical meaning in the Wigner-function formalism.

As an independent check of the Wigner function formalilsm, we perform a calculation of another known QFT object. Substituting the Wigner function (\ref{Y1def}) into the original zilch current (\ref{zilch_tensor_old}) and using (\ref{ord1}) one finds
\bea
&&\left\langle: Z^{(3)3}_{\,\,\,\,\,\,\,\,\,\,\,00}:\right\rangle\bigg|_{r\to0} =-\frac{2}{\hbar^2}\int\,d^4p\,p_{\{0}\left(p_{3\}}U+\e_{3\}\m\n\s}\frac{p^\m n^\n}{p\cdot n}\pa^\s F(\b\cdot p)\right)p_0\,\d(p^2)=\frac{8\p^2T^4}{45\hbar^2}\O \, , \notag
\label{ZVEKarl}
\eea
where we set $U_0=0$ and work in the laboratory frame. Thus the expectation value of the zilch current precisely agrees with the result of \cite{Chernodub:2018era} obtained with the formalism described in Sec.~\ref{Sec:QFT} with the choice $U_0=0$. One also may wonder what is the expectation value of helicity current $K^\m$ and, using the Wigner function (\ref{ord1}), we find
\bea
&&\left\langle: K^i:\right\rangle\bigg|_{r\to0} =\left[-2\int\,d^4p\,\left(2p_i U+\e_{i\n\r\s}\frac{p^\n n^\r}{p\cdot n}\pa^\s F(\b\cdot p)\right)\d(p^2)\right]_{n=n_0}=\frac{4}{9}T^2\O^i \, , \notag
\eea
where $n_0^\m=(1,0,0,0)$ corresponds to the Coulomb gauge. We notice here that the structure in the helicity current differs from the CKT current arising in both definitions of zilch indicating its gauge dependence. We leave the further investigation on the gauge dependence of the photon CVE and its relation to the corrections in the fermionic CVE for future studies.

Finally, in order to clarify the physical meaning of $U_0$, we calculate the zilch charge density of non-rotating photon gases in a grand canonical ensemble. The conservation of zilch as well as helicity allows us to introduce a chiral chemical potential to the distribution function in~(\ref{expect}), namely, $f_B(\x)$ is replaced with $f_B(\x_\l)$ with $\x_\l=\b(\o-\l\m)$. Thus, from~(\ref{zilchsymm}), we obtain
\bea
\left\langle :\bar{Z}^{(s)}_{0..0}:\right\rangle&=&\frac{(-1)^{\frac{s+1}{2}}}{\hbar^{s}}\int\frac{d^3 p}{(2\p)^3}
\o^{s-1}\,\left[f_B(\x_+)-f_B(\x_-)\right]\,,
\eea
where we have restored the $\hbar$ dependence. By comparing with (\ref{wignerz}) and considering the case $\m\ll T$, we find that $U_0=-(\b\m/\hbar)F'(\b\cdot p)$. Thus, $U_0$ represents the effects of a chiral chemical potential of photons. Therefore, our setup with $U_0=0$ corresponds to a photon gas with zero helicity and zilch charge densities at equilibrium.

\section{Summary}

In this work we have studied the chirality transport in a rotating gas of photons and compared its manifestation in the helicity and zilch currents -- the CVE \cite{Avkhadiev:2017fxj} and ZVE \cite{Chernodub:2018era}. We use the freedom in the definition of the general zilch current to introduce a set of higher zilches symmetric in their indices in Sec.~\ref{Sec:QFT}. Using this additional property we identify the corresponding single particle zilch and calculate the full zilch current in a rotating gas of photons within the CKT description introduced in Sec.~\ref{Sec:CKT}. In this framework we find that ZVE is related to the topological Berry phase in the same way as the other chiral effects both in the case of photons and other massless particles with spin. The ZVE contribution to the general zilch current obtained from the semi-classical description (\ref{eq:ZilchCKT}) agrees with an explicit field theory calculation (\ref{eq:ZilchQFT}). We also notice that the universality of the CKT construction \cite{Huang:2018aly}  indicates that the zilch currents can be introduced for particles with an arbitrary spin and there is a family of ZVEs, including ZVE for fermions which has not been discussed in the literature. Furthermore, the relation between the zilch current density in the phase space (\ref{CKTzilchSP}) and the CKT current obtained in this work makes one expect a new contribution to the zilch current -- an anomalous zilch Hall effect - which appears at  second order in the $\hbar$-expansion. One may also expect that the covariant form of (\ref{eq:ZilchCKT}) contains a contribution corresponding to the spin Hall effects which can be thought of as a new zilch spin Hall effect. We leave these new effects for future detailed study elsewhere.

This common origin of ZVE and CVE gives further insight into the relation between the vortical responses in chiral matter and anomalies of the underlying theory. Indeed, if the thermal part of photonic or fermionic CVEs is related to the corresponding mixed gravitational anomaly or global gravitational anomaly, a similar relation should be expected in the case of zilch currents indicating a novel class of anomalies for the zilch currents.

The CKT used in Sec.~\ref{Sec:CKT} is constructed from a single particle semi-classical action, see e.g. \cite{Huang:2018aly}, and the identification of the zilch current is based on its properties under Lorentz transformation. This may seem unsatisfactory if one is interested in how the CKT arises from the microscopic theory. To address this issue in Sec.~\ref{Sec:Wigner} we have constructed the Wigner function for a rotating photon gas for the first time and used it to derive the expectation values of the helicity and zilch currents. We have explicitly seen how the CKT single-particle currents enter in the phase space integrals in the Wigner-function formalism and in this way we support our ``naive'' construction in Sec.~\ref{Sec:CKT}. The Wigner-function formalism also allows us to relate the CVE and ZVE contributions with the Berry phase in the QFT terms.

The Wigner function cannot be solved solely from the EOMs and one has to additionally constraint it from QFT. In (\ref{ord0}) we have used the ansatz for the photon Wigner function in a static uniform gas to fix the leading contribution. At first order in $\hbar$ the Wigner function involves the term (\ref{ord1}) which is transverse both to the momentum $p^\m$ and gauge fixing vector $n^\m$ and is proportional to an unconstrained function $U(x,p)$. However, considering the gauge-invariant Wigner function $Y^{\m\n\r\s}$ allows to fix a part of $U$ required to remove the $n$-dependence from (\ref{Ywithn}). Comparing the results for the zilch currents we have found that the remaining free part $U_0$ should be set to zero for the agreement between the results for ZVE in Sec.~\ref{Sec:QFT} and Sec.~\ref{Sec:Wigner}. Constraining this last term we have obtained the fully defined photon Wigner function (\ref{WF}) and found that ZVE in the original zilch current $Z$
precisely agrees with the results of \cite{Chernodub:2018era}.

Curiously, the intermediate expression for the general symmetrized zilch current shows that the gauge fixing vector $n^\m$ is, in fact, the frame vector introduced in the CKT to fix the definition of the spin tensor. This is not surprising since the full angular momentum of photon cannot be decomposed into orbital and spin parts in a gauge-invariant way. This identification was not presented in the literature before to the best of out knowledge.

The helicity transport of guage fields is especially interesting in light of experimental measurements of hadron spin polarization in off-central heavy-ion collisions at RHIC and LHC~\cite{STAR:2017ckg,Adam:2018ivw,Acharya:2019vpe}.  The final state polarization follows the spin polarization of quarks and gluons in the QGP which, in turn, is in correspondence with the helicity transport due to the chiral effects.
So far, in order to describe the experimental data, most works have considered only spin-half degrees of freedom, see e.g.~\cite{Becattini:2013vja,Becattini:2013fla,Becattini:2015ska,Karpenko:2016jyx,Xie:2017upb,Wei:2018zfb,Xia:2018tes,Liu:2020ymh,Becattini:2020ngo}. However, the contribution of spin-one particles should be also taken into account.
The CVE and ZVE of gauge fields may shed light on their contribution to the spin polarization of QGP and require further investigation. Finally, the Wigner function computed here may serve as a starting point to develop relativstic hydrodynamics with spin degrees of freedom for vector particles~\cite{Florkowski:2017ruc,Florkowski:2017dyn,Florkowski:2018myy,Becattini:2018duy,Florkowski:2018fap,Hattori:2019lfp,Weickgenannt:2020aaf}.

\section*{Acknowledgments}
The authors thank K. Fukushima, Y. Hidaka, K. Landsteiner, D. H. Rischke, N. Weickgenannt, N. Yamamoto, and V.I. Zakharov for enlightening discussions. X.-G.~H. is supported by NSFC through Grants No.~11535012 and No.~11675041. The work of A.V.S. is supported through the LANL/LDRD Program. A.V.S and P.M. are also grateful for support from RFBR Grant 18-02-40056. The work of E.S. is supported by the Deutsche Forschungsgemeinschaft (DFG, German Research Foundation) through the Collaborative Research Center CRC-TR 211 ``Strong-interaction matter under extreme conditions'' -- project number 315477589 - TRR 211. E.S. acknowledges support by BMBF ``Forschungsprojekt: 05P2018 - Ausbau von ALICE am LHC (05P18RFCA1)".

\bibliographystyle{bibstyle}
\bibliography{zilch}

\end{document}